\documentclass{sigchi}

\usepackage{color-edits}
\addauthor{sw}{blue}

\CopyrightYear{2020}
\setcopyright{acmlicensed}
\doi{https://doi.org/10.1145/313831.3376813}
\isbn{978-1-4503-6708-0/20/04}
\conferenceinfo{CHI'20,}{April  25--30, 2020, Honolulu, HI, USA}
\acmPrice{\$15.00}

\toappear{\scriptsize Permission to make digital or hard copies of all or part of this work for personal or classroom use is granted without fee provided that copies are not made or distributed for profit or commercial advantage and that copies bear this notice and the full citation on the first page. Copyrights for components of this work owned by others than ACM must be honored. Abstracting with credit is permitted. To copy otherwise, or republish, to post on servers or to redistribute to lists, requires prior specific permission and/or a fee. Request permissions from permissions@acm.org. \\
{\emph{CHI '20, April 25--30, 2020, Honolulu, HI, USA.} } \\
Copyright is held by the owner/author(s). Publication rights licensed to ACM. \\
ACM ISBN 978-1-4503-6708-0/20/04\ ...\$15.00.\\
http://dx.doi.org/10.1145/313831.3376813}

\usepackage{balance}       
\usepackage{graphics}      
\usepackage[T1]{fontenc}   
\usepackage{txfonts}
\usepackage{mathptmx}
\usepackage[pdflang={en-US},pdftex]{hyperref}
\usepackage{color}
\usepackage{booktabs}
\usepackage{textcomp}
\usepackage{graphicx}
\usepackage{subcaption}
\usepackage{adjustbox}
\usepackage{multirow}
\usepackage{booktabs}
\usepackage{dcolumn}
\usepackage{verbatim}

\usepackage{microtype}        
\usepackage{ccicons}          

\usepackage{todonotes}

\def\plaintitle{SIGCHI Conference Proceedings Format}

\def\emptyauthor{}
\def\plainkeywords{perceived fairness; algorithmic decision-making; algorithm outcome; algorithm development.}

\makeatletter
\def\url@leostyle{%
  \@ifundefined{selectfont}{
    \def\UrlFont{\sf}
  }{
    \def\UrlFont{\small\bf\ttfamily}
  }}
\makeatother
\def\pprw{8.5in}
\def\pprh{11in}

\definecolor{linkColor}{RGB}{6,125,233}
\hypersetup{%
  pdftitle={\plaintitle},
  pdfauthor={\emptyauthor},
  pdfkeywords={\plainkeywords},
  pdfdisplaydoctitle=true, 
  bookmarksnumbered,
  pdfstartview={FitH},
  colorlinks,
  citecolor=black,
  filecolor=black,
  linkcolor=black,
  urlcolor=linkColor,
  breaklinks=true,
  hypertexnames=false
}

\begin{document}

\title{Factors Influencing Perceived Fairness in Algorithmic Decision-Making: Algorithm Outcomes, Development Procedures, and Individual Differences}

\numberofauthors{3}
\author{%
  \alignauthor{Ruotong Wang\\
    \affaddr{Carnegie Mellon University}\\
    \email{ruotongw@andrew.cmu.edu}}\\
  \alignauthor{F. Maxwell Harper\thanks{The work was done while the author was at the University of Minnesota, Twin Cities.}\\
    \affaddr{Amazon}\\
    \email{fmh@amazon.com}}\\
  \alignauthor{Haiyi Zhu\\
    \affaddr{Carnegie Mellon University}\\
    \email{haiyiz@cs.cmu.edu}}\\
}

\maketitle

\begin{abstract}
Algorithmic decision-making systems are increasingly used throughout the public and private sectors to make important decisions or assist humans in making these decisions with real social consequences. While there has been substantial research in recent years to build fair decision-making algorithms, there has been less research seeking to understand the factors that affect people's \textit{perceptions of fairness} in these systems, which we argue is also important for their broader acceptance. In this research, we conduct an online experiment to better understand perceptions of fairness, focusing on three sets of factors: algorithm outcomes, algorithm development and deployment procedures, and individual differences. We find that people rate the algorithm as more fair when the algorithm predicts in their favor, even surpassing the negative effects of describing algorithms that are very biased against particular demographic groups. We find that this effect is moderated by several variables, including participants' education level, gender, and several aspects of the development procedure. Our findings suggest that systems that evaluate algorithmic fairness through users' feedback must consider the possibility of  ``outcome favorability'' bias.
\end{abstract}




\section{Introduction}

Algorithmic systems are widely used in both the public and private sectors for making decisions with real consequences on people's lives. For example, ranking algorithms are used to automatically determine the risk of undocumented immigrants to public safety~\cite{kalhan_immigration_2013}. Filtering algorithms are used in job hiring and college admission processes~\cite{ajunwa_hiring_2016, kuncel_hiring_2014}. Scoring algorithms are used in loan approval~\cite{odwyer_algorithms_nodate}.

While algorithms have the potential to make decision-making more efficient and reliable~\cite{cowgill_bias_nodate,erel_research:_2018,hurley_can_2018,kleinberg_human_2017,miller_want_2018}, concerns about their fairness may prevent them from being broadly accepted~\cite{noauthor_nsf_nodate}. In one case, Amazon abandoned an algorithmic recruitment system for reviewing and ranking applicants' resumes because the system was biased against women~\cite{dastin_amazon_2018}. In another case, an algorithm for predicting juvenile delinquency in St. Paul, Minnesota was derailed due to public outcry from the community over concerns about bias against children of color~\cite{pomeroy_how_nodate}. 

There has been an increasing focus in the research community on understanding and improving the fairness of algorithmic decision-making systems. For example, fairness-aware (or discrimination-aware) machine learning research attempts to translate fairness notions into formal algorithmic constraints and develop algorithms subject to such constraints (e.g.,~\cite{berk2018fairness, chouldechova_frontiers_2018, corbett-davies_algorithmic_2017, hardt2016equality, kleinberg2016inherent,lepri_fair_2018}) However, there is a disconnect between the theoretical discrimination-aware machine learning approaches and the behavioral studies investigating how people perceive fairness of algorithms that affect their lives in real-world contexts (e.g., ~\cite{lee_human-centered_2017,veale_fairness_2018}). People's perception of fairness can be complicated and nuanced. For example, one interview study found that different stakeholders could have different notions of fairness regarding the same algorithmic decision-making system~\cite{lee_human-centered_2017}. Other studies suggested disagreements in people's fairness judgements~\cite{grgic2018human, srivastava2019mathematical}. Our research attempts to connect the two lines of literature by asking how the the theoretical fairness notion translates into perceptions of fairness in practical scenarios, and how other factors also influence people's perception of fairness.

In this research, we systematically study what factors influence people's perception of the fairness of algorithmic decision-making processes. The first set of factors we investigate is  \textbf{\textit{algorithm outcomes}}. Specifically, we explore both the individual-level and group-level outcomes: whether the algorithm is favorable or unfavorable to specific \textit{individuals}, and whether the algorithm is biased or unbiased against specific \textit{groups}. Specifically, the biased outcome is operationalized by showing high error rates in protected groups, while the unbiased outcome is showing very similar error rates across different groups. Note that this operationalization is directly aligned with the prevalent theoretical fairness notion in the fairness-aware machine learning literature, which asks for approximate equality of certain statistics of the predictor, such as false positive rates and false-negative rates, across different groups.

In addition to \textbf{\textit{algorithm outcomes}} (i.e., whether an algorithm's prediction or decision is favorable to specific individuals or groups), we also investigate \textbf{\textit{development procedures}} (i.e., how an algorithm is created and deployed), and \textbf{\textit{individual differences}} (i.e., the education or demographics of the individual who is evaluating the algorithm). Specifically, we investigate the following research questions:
\begin{itemize}
    \item How do (un)favorable outcomes to individuals and (un)biased treatments across groups affect the perceived fairness of algorithmic decision-making? 
    \item How do different approaches to algorithm creation and deployment (e.g., different levels of human involvement) affect the perceived fairness of algorithmic decision-making?
    \item How do individual differences (e.g., gender, age, race, education, and computer literacy) affect the perceived fairness of algorithmic decision-making?
\end{itemize}

To answer these questions, we conducted a randomized online experiment on Amazon Mechanical Turk (MTurk). We showed participants a scenario, stating ``Mechanical Turk is experimenting with a new algorithm for determining which workers earn for a Masters Qualification.'' \footnote{``Masters'' workers can access exclusive tasks that are often associated with higher payments~\cite{matsakis_unknown_2016}).} We asked MTurk workers to judge the fairness of an algorithm that the workers can personally relate to. MTurk workers are stakeholders in the problem, and their reactions are representative of laypeople who are affected by the algorithm’s decisions.
    
In the description we presented to the participants, we included a summary of error rates across demographic groups and a description of the algorithm's development process, manipulating these variables to test different plausible scenarios. We included manipulation check questions to ensure that the participants understood how the algorithm works. We showed each participant a randomly chosen algorithm output (either ``pass'' or ``fail''), to manipulate whether the outcome would be personally favorable or not. Participants then answered several questions to report their perception of the fairness of this algorithmic decision-making process. We concluded the study with a debriefing statement, to ensure that participants understood this was a hypothetical scenario, and the algorithmic decision was randomly generated.

We found that perceptions of fairness are strongly increased both by a  \textit{favorable outcome} to the individual (operationalized by a ``pass'' decision for the master qualification), and by the \textit{absence of biases} (operationalized by showing very similar error rates across different demographic groups). The effect of a favorable outcome at individual level is larger than the effect of the absence of bias at group level, suggesting that solely satisfying the statistical fairness criterion does not guarantee perceived fairness. Moreover, the effect of a favorable or unfavorable outcome on fairness perceptions is mitigated by additional years of education, while the negative effect of including biases across groups is exacerbated by describing a development procedure with ``outsourcing'' or a higher level of transparency. Overall, our findings point to the complexity of understanding perceptions of fairness of algorithmic systems.

\section{Related Work and Hypothesis}

There is a growing body of work that aims to improve fairness, accountability, and interpretability of algorithms, especially in the context of machine learning. For example, much fairness-aware machine learning research aims to build predictive models that satisfy fairness notions formalized as algorithmic constraints, including statistical parity~\cite{dwork_fairness_2012}, equalized opportunity~\cite{hardt2016equality}, and calibration~\cite{kleinberg_human_2017}. For many of these fairness measures there are algorithms that explore trade-offs between fairness and accuracy~\cite{agarwal_reductions_2018, berk_fairness_2018,dwork_decoupled_2018, menon_cost_2018}. For interpreting a trained machine learning model, there are three main techniques: sensitivity or gradient-based analysis~\cite{ribeiro_why_2016,koh_understanding_2017}, building mimic models~\cite{hendricks_generating_2016}, and investigating hidden layers~\cite{bau_network_2017, veale_fairness_2018}. However, Veale et al. found that these approaches and tools are often built in isolation of specific users and user context~\cite{veale_fairness_2018}. HCI researchers have conducted surveys, interviews, and analyses on public tweets to understand how real-world users perceive and adapt to algorithmic systems~\cite{devito_algorithms_2017, devito_too_2018,eslami_first_2016, french_whats_2017, lee_algorithmic_2017, lee_working_2015}. However, to our knowledge, the perceived fairness of algorithmic decision-making has not been systematically studied. 

\subsection{Human Versus Algorithmic Decision-making}

Social scientists have long studied the sense of fairness in the context of human decision-making. One key question is whether we can apply the rich literature on the fairness of \textit{human} decision-making to the fairness of \textit{algorithmic} decision making. 

On one hand, it is a fundamental truth that an algorithm is not a person and does not warrant human treatment or attribution~\cite{nass_machines_2000}. Prior work shows that people treat these two types of decision-making differently (e.g., ~\cite{dietvorst_algorithm_nodate,lee_understanding_2018,shank_perceived_2012}). For example, researchers describe ``algorithm aversion,'' where people tend to trust humans more than algorithms even when the algorithm makes more accurate predictions. This is because people tend to quickly lose confidence in a algorithm after seeing that it makes mistakes~\cite{dietvorst_algorithm_nodate}. Another study pointed out that people attribute fairness differently: while human managers' fairness and trustworthiness were evaluated based on the person's authority, algorithms' fairness and trustworthiness were evaluated based on efficiency and objectivity~\cite{lee_understanding_2018}. On the other hand, a series of experimental studies demonstrated that humans mindlessly apply social rules and expectations to computers~\cite{nass_machines_2000}. One possible explanation is that the human brain has not evolved quickly enough to assimilate the fast development of computer technologies~\cite{reeves_media_1996}. Therefore, it is possible that research on human decision-making can provide insights on how people will evaluate algorithmic decision-making.

In this paper, we examine the degree to which several factors affect people's perception of fairness of algorithmic decision-making. We study factors identified by the literature on fairness of human decision-making --- algorithm outcomes, development procedures, and interpersonal differences --- as we believe that they may also apply in the context of algorithmic decision-making.

\subsection{Effects of Algorithm Outcomes}

When people receive decisions from an algorithm, they will see the decisions as more or less favorable to themselves, and more or less fair to others. Meta-review by Skitka et al. revealed that outcome favorability is empirically distinguishable from outcome fairness \cite{skitka_are_2003}.  

Our first hypothesis seeks to explore the effects of outcome favorability on perceived fairness. Psychologists and economists have studied perceived fairness  extensively, particularly in the context of resource allocation (e.g.,~\cite{batson_two_2016,diekmann_self-interest_1997,major_chapter_1982, tyler_social_2000}). They found that the perception of fairness will increase when individual receive outcomes that are favorable to them (e.g.,~\cite{tyler_social_2000}). A recent survey study on 209 litigants showed that litigates who receive favorable outcomes (e.g., a judge approves their request) will perceive their court officials to be more fair and will have more positive emotions towards court officials~\cite{hou_factors_2017}. Outcome favorability is also associated with fairness-related consequences. Meta-analysis shows that outcome favorability explains 7\% of the variance in organizational commitment and 24\% of the variance in task satisfaction~\cite{skitka_are_2003}.   

{\it {\bf Hypothesis 1a:} People who receive a favorable outcome think the algorithm is more fair than people who receive an unfavorable outcome.}

Social scientists suggest that people judge outcome fairness by ``whether the proportion of their inputs to their outcomes are comparable to the input/outcome ratios of others involved in the social exchange''~\cite{skitka_are_2003}, often referred to as ``equity theory'' or ``distributive justice''~\cite{adams_inequity_1965, homans_social_1974, walster_equity:_1978}. 
Contemporary theorists suggest that distributive justice is influenced by which distributive norm people use in the relational context (e.g., ~\cite{m_distributive_1985, lerner_justice_1974}), the goal orientation of the allocator (e.g.,~\cite{m_distributive_1985}), the resources being allocated (e.g.,~\cite{skitka_ideological_1999}), and sometimes political orientation (e.g.,~\cite{major_chapter_1982}). 
Empirical studies show that a college admission process is perceived as more fair if it does not consider gender and if it is not biased against any gender groups ~\cite{nacoste_but_1987}. Another study show that an organization will be perceived as less fair when managers treat different employees differently~\cite{lamertz_social_2002}.

In the context of algorithmic decision-making, one common way of operationalizing equity and outcome fairness is ``overall accuracy equality'', which considers the disparity of accuracy between different subgroups (e.g., different gender or racial groups)~\cite{berk2018fairness}. The lower the accuracy disparity between different groups, the less biased the algorithm is. We hypothesize that people will judge an algorithm to be more fair when they know it is not biased against any particular subgroups. 

{\it {\bf Hypothesis 1b:} People perceive algorithms that are not biased against particular subgroups as more fair.}

Prior literature provides mixed predictions regarding the relative effects of outcome favorability (i.e., whether individuals receive favorable outcome or not) and outcome fairness (i.e., whether different groups receive fair and unbiased treatments) on the perceived fairness of the algorithm. A majority of the researchers believe that individuals prioritize self-interest over fairness for everyone. Economists believe that people are in general motivated by self-interest and are relatively less sensitive towards group fairness~\cite{rodriguez-lara_self-interest_2012,rutstrom_entitlements_2000}. Specifically, individuals are not willing to sacrifice their own benefits to pursue a group common good in resource redistribution tasks~\cite{esarey_what_2012}. A similar pattern has been found in the workplace. A survey of hotel workers showed that people displayed the highest level of engagement and the lowest rate of burnout when they were over-benefited in their work, receiving more than they think they deserved. In addition, people tend to justify an unequal distribution when they are favored in that distribution \cite{moliner_perceived_2013}.

On the other hand, research also shows that people prioritize fairness over personal benefits in certain situations. For example, bargainers might be reluctant to benefit themselves when it harms the outcomes of others, contingent on their social value orientations, the valence of outcomes, and the setting where they negotiate \cite{beest_self-interest_2007}. In an analysis of presidential vote choices, researchers found that voters are more likely to vote for the president who will treat different demographic subgroups  equally, independent of voters' own group membership \cite{mutz_dimensions_1997}.
Whether the decision takes place in public or in private has a strong impact. In a study conducted by Badson et al., participants had to choose between allocation of resources to the group as a whole or to themselves alone. When the decision was public, the proportion allocated to the group was 75\%. However, once the decisions became private, the allocation to the group dropped to 30\% \cite{batson_two_2016}.

Algorithmic decisions are often private, not under the public scrutiny. Therefore, we hypothesize that people will prioritize  self-interest when they react to the algorithm's decisions. As a result, the effect of a favorable outcome will be stronger than the effect of an unbiased treatment. 

{\it {\bf Hypothesis 1c:} In the context of algorithmic decision-making, the effect of a favorable outcome on perceived fairness is larger than the effect of being not biased against particular groups.}

\subsection{Effects of Algorithm Development Procedures}

Procedural fairness theories concentrate on the perceived fairness of the procedure used to make decisions (e.g., ~\cite{folger1985procedural}). For example, Gilliland examined the procedural fairness of the employment-selection system in terms of ten procedural rules, including job relatedness, opportunity to perform, reconsideration opportunity, consistency of administration, feedback, selection information, honesty, interpersonal effectiveness of administrator, two-way communication, propriety of questions and equality needs ~\cite{gilliland_perceived_1993}. 

Specifically, transparency of the decision-making process has an important impact on the perceived procedural fairness. For example, a longitudinal analysis highlighted the importance of receiving an explanation from the organization about how and why layoffs were conducted, in measuring the perceived fairness of layoff decisions~\cite{wanberg_perceived_1999}. 

In the context of algorithmic decision-making, transparency also influences people's perceived fairness of the algorithms. Researchers have shown that some level of transparency of an algorithm would increase users' trust of the system even when the user's expectation is not aligned with the algorithm's outcome~\cite{kizilcec_how_2016}. However, providing too much information about the algorithm might have the risk of confusing people, and therefore reduce the perceived fairness~\cite{kizilcec_how_2016}.

{\it {\bf Hypothesis 2a: } An algorithmic decision-making process that is more transparent is perceived as more fair than a process that is less transparent.}

The level of human involvement in the creation and deployment of an algorithm might also play an important role in determining perceptions of fairness. Humans have their own bias. For example, sociology research shows that hiring managers tend to favor candidates that are culturally similar to themselves, making hiring more than an objective process of skill sorting~\cite{rivera_hiring_2012}.

However, in the context of algorithmic decision-making, human involvement is often viewed as a mechanism for correcting machine bias. This is especially the case when people become increasingly aware of the limitation of algorithms in marking subjective decisions. In tasks that require human skills such as hiring and work evaluation, human decisions are perceived as more fair, because algorithms are perceived as lacking intuition and the ability to make subjective judgments~\cite{lee_understanding_2018}. The aversion to algorithmic decision-making could be mitigated when users contribute to the modification of the algorithms~\cite{dietvorst_overcoming_2016}. Recent research on the development of decision-making algorithms has advocated for a human-in-the-loop approach, in order to make the algorithm more accountable~\cite{ito_society_2016, rahwan_society---loop:_2018, schirner_future_2013, zhu_value-sensitive_2018}. In sum, human involvement is often considered as a positive influence in decision-making systems, due to the human's ability to recognize factors that are hard to quantify. Thus, it is reasonable to hypothesize that people will evaluate algorithms as more fair when there are human insights involved in the different steps of the algorithm creation and deployment. 

{\it {\bf Hypothesis 2b:} An algorithmic decision-making process that has more human involvement is perceived as more fair than a process that has less human involvement.}

\subsection{Effects of Individual Differences}
In this research, we also investigate the extent to which the perceptions of algorithmic fairness are influenced by people's personal characteristics. Specifically, we look at two potential influential factors: education and demographics.

We first consider education. We believe both general education and computer science education may influence perceptions of fairness. People with greater knowledge in general and about computers specifically may simply have a better sense for what types of information an algorithm might process, and how that information might be processed to come to a decision. We are not aware of research that has investigated the link between computer literacy and perceptions of fairness in algorithmic decision-making, but research has investigated related issues. For example, prior work has looked at how people form ``folk theories'' of algorithms' operation, often leading to incorrect notions of their operation and highly negative feelings about the impact on their self interest ~\cite{devito_algorithms_2017}. We speculate that users with greater computer literacy will have more realistic expectations, leading them to more often agree with the perspective that an algorithm designed to make decisions is likely to be a fair process. 

Pierson, et al. conducted a survey of undergraduate students before and after an hour-long lecture and discussion on algorithmic fairness, finding that students' views changed; in particular, more students came to support the idea of using algorithms rather than judges (who might themselves be inaccurate or biased) in criminal justice~\cite{pierson_demographics_2017}. This finding suggests that education, particularly education to improve algorithmic literacy, may lead to a greater perception of algorithmic fairness.

{\it {\bf Hypothesis 3a:} People with a higher level of education will perceive algorithmic decision-making to be more fair than people with a lower level of education.}

{\it {\bf Hypothesis 3b:} People with high computer literacy will perceive algorithmic decision-making to be more fair than people with low computer literacy.}

It is possible that different demographic groups have different beliefs concerning algorithmic fairness. Research has found that privileged groups are less likely to perceive problems with the status quo; for example, the state of racial equality is perceived differently by whites and blacks in the United States~\cite{pew_research_center_views_2016}. 

There is a growing body of examples documenting algorithmic bias against particular groups of people. For example, an analysis by Pro Publica found that an algorithm for predicting each defendant's risk of committing future crime was twice as likely to wrongly label black defendants as future criminals, as compared with white defendants~\cite{angwin_machine_2016}. In another case, an algorithmic tool for rating job applicants at Amazon penalized resumes containing the word ``women's''~\cite{dastin_amazon_2018}. 

There has been little work directly investigating the link between demographic factors and perceptions of algorithmic fairness. One recent study did not find differences between men and women in their perceptions of fairness in an algorithmic decision-making process, but did find that men were more likely than women to prefer maximizing accuracy over minimizing racial disparities in a survey describing a hypothetical criminal risk prediction algorithm~\cite{pierson_demographics_2017}.

Based on the growing body of examples of algorithmic bias against certain populations, we predict that different demographic groups will perceive fairness differently:

{\it {\bf Hypothesis 3c:} People in demographic groups that typically benefit from algorithmic biases (young, white, men) will perceive algorithmic decision-making to be more fair than people in other demographic groups (old, non-white, women).}

\section{Method}

To test the hypotheses about factors influencing people's perceived fairness of algorithmic decision-making, we conducted a randomized between-subjects experiment on MTurk. The context for this experiment is a description of an ``experimental'' algorithmic decision-making process that determines which MTurk workers are awarded a Master qualification.

\subsection{Study platform}
MTurk is an online crowdsourcing workplace where over 500,000 workers complete microtasks and get paid by requesters~\cite{pew_research_center_mechanical_2016}. Masters workers are the ``elite groups of workers who have demonstrated accuracy on specific types of HITs on the Mechanical Turk marketplace''~\cite{noauthor_amazon_nodate}. Since Masters workers can access exclusive tasks that are usually associated with higher payments, the Master qualification is desirable among MTurk workers~\cite{matsakis_unknown_2016}. 

The Master qualification is a black-box process to workers. According to Amazon's FAQs page, Mechanical Turk uses ``statistical models that analyze Worker performance based on several Requester-provided and marketplace data points'' to decide which workers are qualified~\cite{noauthor_amazon_nodate}.

\subsection{Experiment Design }

\begin{table}[h]
\huge
\begin{adjustbox}{width=1\columnwidth}
\begin{tabular}{|l|l|l|}
\hline
\textbf{Manipulations}                & \textbf{Conditions} & \textbf{Descriptions shown to participants}                                                                                                                                                                                                                                          \\ \hline
\multicolumn{3}{|c|}{\textbf{Algorithm Outcome (H1)}}                                                                                                                                                                                                                                                                                              \\ \hline
\multirow{2}{*}{\begin{tabular}[c]{@{}l@{}}(Un)favorable \\Outcome\end{tabular} } & Favorable           & \textit{\begin{tabular}[c]{@{}l@{}} The algorithm has processed your HIT history,\\ and the result is positive (you passed the Master qualification test).\end{tabular}}                                                                                                                        \\ \cline{2-3} 
                                      & Unfavorable         & \textit{\begin{tabular}[c]{@{}l@{}}The algorithm has processed your HIT history,\\ and the result is negative (you did not pass the Master qualification test).\end{tabular}}                                                                                                                  \\ \hline
\multirow{2}{*}{\begin{tabular}[c]{@{}l@{}}(Un)biased \\Treatment\end{tabular} } & Biased              & \textit{\begin{tabular}[c]{@{}l@{}}Percent errors by gender: Male 2.6\%, Female 10.7\%\\ Percent errors by age: Above 45 9.8\%, Between 25 and 45 3.6\%, Below 25 1.2\%\\ Percent errors by race: White 0.7\%, Asian 6.4\%, Hispanic 5.8\%, \\Native American 6.7\%, Other 14.1\%\end{tabular}} \\ \cline{2-3} 
                                      & Unbiased            & \textit{\begin{tabular}[c]{@{}l@{}}Percent errors by gender: Male 6.4\%, Female 6.3\%\\ Percent errors by age: Above 45 6.4\%, Between 25 and 45 6.2\%, Below 25 6.5\%\\ Percent errors by race: White 6.5\%, Asian 6.4\%, Hispanic 6.4\%, \\Native American 6.5\%, Other 6.2\%\end{tabular} }  \\ \hline
\multicolumn{3}{|c|}{\textbf{Algorithm Creation and Deployment (H2)}}                                                                                                                                                                                                                                                                              \\ \hline
\multirow{2}{*}{Transparency}         & High                &\textit{ \begin{tabular}[c]{@{}l@{}}The organization publishes many aspects of this computer algorithm on the web, \\ including the features used, the optimization process, and benchmarks \\ describing the process's accuracy.\end{tabular} }                                                \\ \cline{2-3} 
                                      & Low                 & \textit{\begin{tabular}[c]{@{}l@{}}The organization does not publicly provide any information \\ about this computer algorithm.\end{tabular}  }                                                                                                                                                                                            \\ \hline
\multirow{3}{*}{Design}               & CS Team             &\textit{ \begin{tabular}[c]{@{}l@{}}The algorithm was built and optimized by a team of computer scientists \\within the organization.  \end{tabular}}                                                                                                                                                                                     \\ \cline{2-3} 
                                      & Outsourced          & \textit{\begin{tabular}[c]{@{}l@{}}The algorithm was outsourced to a company that specializes \\in applicant assessment software    \end{tabular} }                                                                                                                                                                                       \\ \cline{2-3} 
                                      & CS and HR           & \textit{\begin{tabular}[c]{@{}l@{}}The algorithm was built and optimized by a team of computer scientists \\ and other MTurk staff across the organization.\end{tabular}}                                                                                                                     \\ \hline
\multirow{2}{*}{Model}                & \begin{tabular}[c]{@{}l@{}}Machine \\Learning\end{tabular}     & \textit{\begin{tabular}[c]{@{}l@{}}The algorithm is based on machine learning techniques trained to recognize \\ patterns in a very large dataset of data collected by the organization over time\end{tabular} }                                                                              \\ \cline{2-3} 
                                      & Rules               & \textit{\begin{tabular}[c]{@{}l@{}}The algorithm has been hand-coded with rules provided by \\ domain experts within the organization.                               \end{tabular}}                                                                                                                                                    \\ \hline
\multirow{2}{*}{Decision}             & Mixed               & \textit{\begin{tabular}[c]{@{}l@{}}The algorithm's decision is then considered on a case-by-case basis by \\ MTurk staff within the organization.                                                      \end{tabular} }                                                                                                                                                 \\ \cline{2-3} 
                                      & \begin{tabular}[c]{@{}l@{}}Algorithm\\-only\end{tabular}       & \textit{The algorithm is used to make the final decision.                                }                                                                                                                                                                                                    \\ \hline
\end{tabular}
\end{adjustbox}
\caption{\textbf{Summary of the experimental manipulations shown to participants.}}
\label{tbl:experimentdesign}
\end{table}

We designed an online between-subjects experiment in which participants were randomly assigned into a 2 (biased vs. unbiased treatment to groups) {$\times$} 2 (favorable vs. unfavorable outcome to individuals) {$\times$} 2 (high vs. low transparency) {$\times$} 3 (outsourced vs. computer scientists vs. mixed design team) {$\times$} 2 (machine learning vs. expert rule-based model)  {$\times$} 2 (algorithm-only vs. mixed decision) design. These manipulations allow us to test the effects of algorithm outcome on perceived fairness (Hypothesis 1) and the effects of algorithm development procedures on perceived fairness (Hypothesis 2). We asked participants to provide their demographic information, which allows us to test Hypothesis 3. The experiment design and manipulations are summarized in Table~\ref{tbl:experimentdesign}.

\begin{figure}[h]
    \centering
    \includegraphics[width=1\columnwidth]{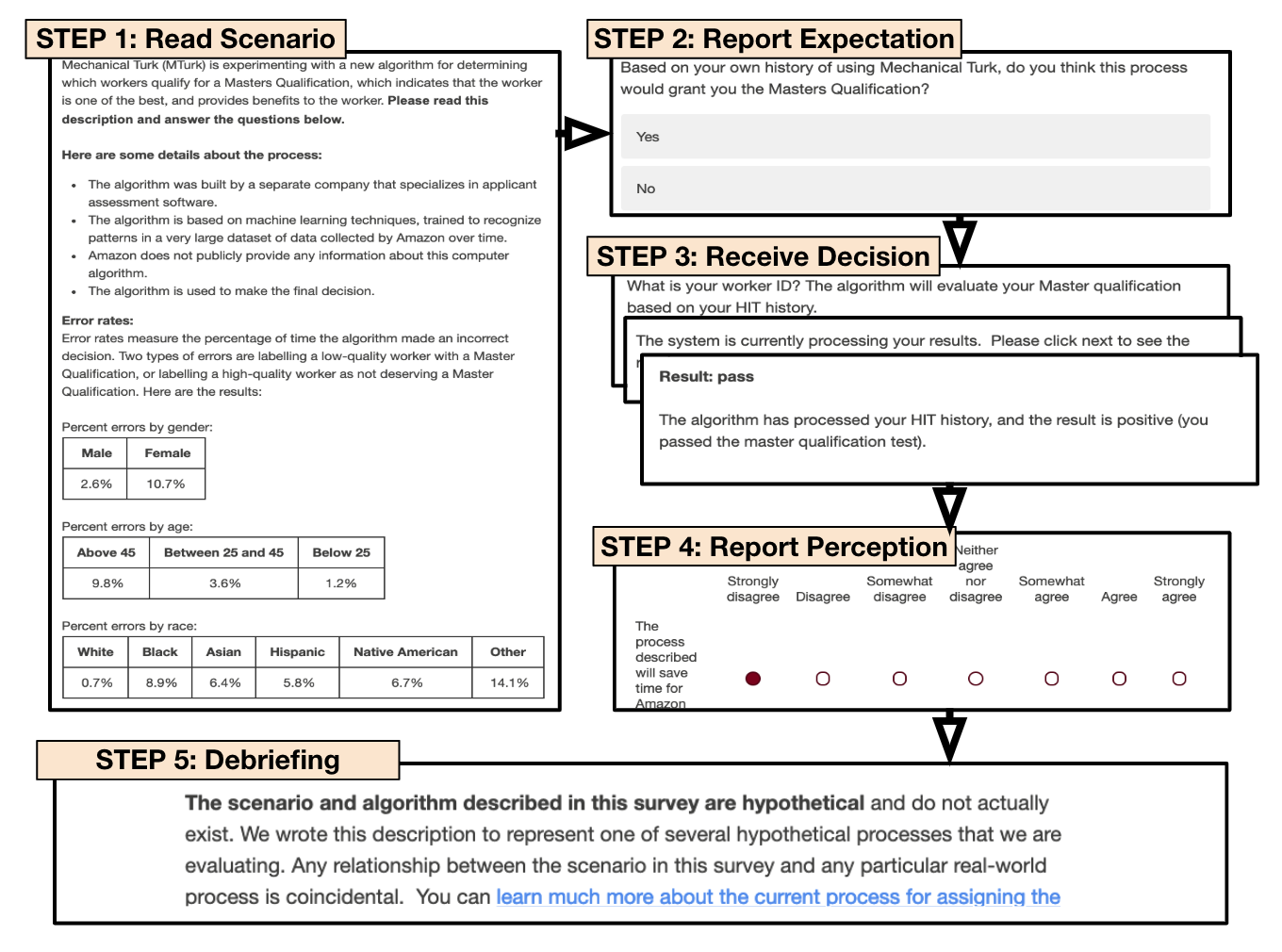}
    \caption{\textbf{Overview of the procedure of the experiment for each participant.}
    }
    \label{fig:screenshot}
\end{figure}

{\bf Algorithm Outcome:} We explore two aspects of algorithm outcome, (un)favorable outcome to individual and (un)biased treatment to group.
    \begin{itemize}
         \item {\bf (Un)favorable outcome:} Participants randomly received a decision and were told the decision was was generated by the algorithm. The decision (``pass'' or ``fail'') corresponds to either a favorable or an unfavorable outcome. 
         \item {\bf (Un)biased treatment:} In both conditions, participants were shown tables of error rates across demographic groups (see Figure~\ref{fig:screenshot}). In the unbiased condition, participants saw very similar error rates across different demographic groups, which (approximately) satisfies the fairness criterion of “overall accuracy equality” proposed by Berk et al.~\cite{berk2018fairness}; in the biased condition, participants saw different error rates across different groups (we used error rates from a real computer vision algorithm from a previous study~\cite{buolamwini_gender_2018}), which violates the same statistical fairness criterion. 
    \end{itemize}

{\bf Algorithm Development Procedure:} We also manipulated the transparency and level of human involvement in the algorithm creation and deployment procedure. 
Participants were shown a description of the algorithm corresponding to their randomly-assigned conditions. The specific text of each condition is shown in Table~\ref{tbl:experimentdesign}.

\subsection{Procedure}

The experiment consisted of five steps, as shown in Figure~\ref{fig:screenshot}. 

\textbf{Step 1}. Participants were shown information about the algorithm in two parts. The first part provided details about the algorithm development and design process. We created 24 (3$\times$2$\times$2$\times$2) different variations for this part, manipulating the four aspects of algorithm development (transparency, design, model, and decision). The second part of the description  is a table of error rates across different demographic subgroups. The error rates are either consistent or differentiated across groups. Participants were required to answer quiz questions to ensure that they paid attention to the description. If they participants gave the incorrect answer, they could go back to read the description again. However, they could not proceed to the next step until they answered all the quiz questions correctly. 

\textbf{Step 2}. We asked participants to self-evaluate whether they think the algorithm will give them a pass or fail result for a Master qualification. 

\textbf{Step 3}. We asked participants to submit their worker ID. We explained that our algorithm has a way to retrieve their HIT history and that the algorithm will evaluate their Master qualification based on their prior performance. Then, we showed participants their outcome: pass or fail. While we told participants that this result was generated by the algorithm, it was, in fact, randomly generated (we did not feed their worker ID and HIT history to an algorithm).

\textbf{Step 4}. We asked participants to answer a set of survey questions related to the fairness of the algorithm and collected their demographic information. We also included the following attention check question: ``This is an attention check. Choose Disagree''. 
At the end of the survey, participants were given the opportunity to freely express their opinions regarding the algorithmic decision-making process in an open-ended question.

\textbf{Step 5}. We concluded the survey with a debriefing statement, where we clarified the deceptions in our study design. We clearly stated that (a) the description about the MTurk Master qualification algorithm is made-up, both in terms of the algorithm error rate and the development and deployment process and (b) the result of the Master qualification algorithm was randomly generated.

\subsection{Operationalization:}

\subsubsection{Dependent variables:} 
\begin{itemize}
    \item {\bf Perceived fairness:} Participants rated their fairness perceptions towards the decision-making algorithm after they learned the design process and experienced the outcome. Their perceived fairness is measured by six questions adapted from previous studies~\cite{franke_does_2012,lee_algorithmic_2017}. All questions are on a 7-point likert scale, from ``strongly disagree'' to ``strongly agree''. The questions included ``the process described is fair'', ``the process described is fair to job-seekers'', ``the process described is fair to the employer organization'', ``the decisions that the organization makes as a result of this process will be fair'', ``the process described will lead the organization to make great hiring decisions'', and ``the process described will make mistakes''. The mean for perceived fairness in our data is 4.0, and the median is 4.3. The internal consistency (Cronbach's alpha) of the six questions is 0.91, which is above the ``excellent'' standard of 0.9.
\end{itemize}

\subsubsection{Independent variables:}
\begin{itemize}
    \item {\bf Algorithm Outcomes:}
        \begin{itemize}
            \item {\it (Un)favorable Outcome:} This variable takes the value of 1 for participants who receive the ``unfavorable'' outcome, and 0 for ``favorable''.
            \item {\it (Un)biased Treatment:} This variable takes the value of 1 for participants who viewed ``biased'' error rates across groups, and 0 for ``unbiased''.
        \end{itemize}
    \item {\bf Algorithm Design and development:}
        \begin{itemize}
            \item {\it Transparency:} This variable takes the value of 1 for participants who viewed the ``high'' transparency description, and 0 for ``low''.
            \item {\it Design:} We operationalize the three types of design (outsourced, cs, cs and hr) as two dummy variables. {\it CS team} and {\it CS and HR}, which take the value of 1 for participants in that respective condition. 
            \item {\it Model:} This variable takes the value of 1 for participants who viewed the ``machine learning'' description, and 0 for ``rules''.
            \item {\it Decision:} This variable takes the value of 1 for participants who viewed ``mixed'', and 0 for ``algorithm-only''.
        \end{itemize}
    \item {\bf Individual differences:} 
        \begin{itemize}
            \item {\it Degree:} We asked participants to report their  highest completed degrees. We grouped the participants into three categories, ``above Bachelor's degree'', ``Bachelor's degree'', and ``below Bachelor's degree''. We use ``above Bachelor's degree'' as a baseline in the analysis. 
            \item {\it Computer Literacy:} We developed 8 questions to evaluate participants' computer literacy, which assessed participants' computer skills, familiarity, and knowledge. Six questions are on a 7-point Likert scale, and two are on a 4-point scale. We normalized the 4-point scale answers into 7-point scale and composited them to calculate the final literacy score. The three questions on computer skills 
            are adapted from \cite{tsai_developing_2018} and \cite{lee_algorithmic_2017}. The two questions on computer familiarity 
            are adapted from \cite{wilkinson_construction_2010}. Since our study is specifically about decision-making algorithms, which is different from previous studies measuring computer literacy, we developed three new questions focusing on participants' knowledge of algorithms. 
            The original questionnaires are included in the supplementary material. 
            The internal consistency (Cronbach's alpha) for all the eight questions is 0.74, which is above the ``acceptable'' standard. The mean for computer literacy in our data is 4.94 and the median is 5. In our analysis, we grouped the numeric computer literacy factor into a binary categorical variable where scores lower than or equal to 5 are labelled ``Low Literacy'' and otherwise ``High Literacy'' (the baseline). 
            \item {\it Age:} We grouped the age data provided by participants into three categories, ``below 25'', ``between 25--45'', and ``above 45''. We use ``Between 25--45'' as a baseline to create two dummy variables on age. 
            \item {\it Gender:} We provided an optional text input box at the end of survey to allow participants to indicate their preferred gender. In our analysis, we only included data from participants who identify themselves as either male or female. 
            \item {\it Race:} We allowed participants to indicate all race categories that apply to them. We then created a dummy variable for each racial group, which resulted in five variables. For each variable, belonging to the racial group is labelled as 1 and 0 otherwise.
        \end{itemize}
\end{itemize}

\subsubsection{Control Variables:}
\begin{itemize}
    \item {\bf Self-expectation:} This variable takes the value of 1 for participants who expected to pass the Master qualification before receiving the algorithm outcome, and 0 for participants who expected to fail.
\end{itemize}

\subsection{Participants}
We recruited 590 participants from MTurk in January and February 2019. All participants are based in the United States and have overall HIT approval rates of at least 80\%. Each participant received \$1.50 as a compensation after finishing the experiment. We excluded 11 participants who failed the attention check question in our data analysis, leaving 579 responses. 153 participants reported a self-expectation of ``fail'', while the remaining 426 answered ``pass''. The random algorithm gave 292 passes and 287 fails; 287 participants received an outcome that aligned with their self-expectation while 292 received decisions that did not align. 

\section{Results}

\subsection{Descriptive Statistics and Statistical Analysis}

\begin{table}[h]
\begin{adjustbox}{width=\columnwidth}
\begin{tabular}{lccccccc}
\\[-1.8ex]\hline 
\hline \\[-1.8ex]
\multicolumn{8}{c}{\textbf{Categorical Variables}}                                                                                                                                                                                                                \\ \cline{1-8} 
                                       & Age                     &                         & Gender                  &                         & Degree                  &                         & \begin{tabular}[c]{@{}c@{}}Computer \\ Literacy\end{tabular} \\ \hline
Above 45                               & 423                     & Female                  & 205                     & Above                   & 70                      & High                    & 264                                                          \\
Btw 25-45                              & 107                     & Male                    & 335                     & Bachelor                & 247                     & Low                     & 315                                                          \\
Below 25                               & 49                      & Other                   & 39                      & Below                   & 262                     &                         &                                                              \\ \hline 
\hline \\ [-2ex]
\multicolumn{8}{c}{\textbf{Continuous Variable}}                                                                                                                                                                                                                  \\ \cline{2-8} 
                                       & Min                     & 1st Qu.                 & Median                  & Mean                    & 3rd Qu.                 & Max                     & S.D.                                                         \\ \hline
\multicolumn{1}{l}{Perceived Fairness} & \multicolumn{1}{l}{1.0} & \multicolumn{1}{l}{2.8} & \multicolumn{1}{l}{4.3} & \multicolumn{1}{l}{4.0} & \multicolumn{1}{l}{5.2} & \multicolumn{1}{l}{7.0} & \multicolumn{1}{l}{1.5} \\                        \hline 
\hline \\[-1.8ex] 
\end{tabular}
\end{adjustbox}
\caption{\textbf{Descriptive statistics of the 579 responses}}
\label{tbl:descriptive}
\end{table}

The descriptive statistics of major variables are shown in Table  \ref{tbl:descriptive}. We used linear regression models to analyze the data and test our hypotheses, with perceived fairness as the dependent variable, and algorithm outcomes, development procedures, and interpersonal differences as the independent variables. We also controlled for the self-assessment of whether they will pass or not. We report coefficients, p-values, standard errors, {$R^2$} and adjusted {$R^2$} values.

\subsection{(Un)favorable outcome vs. (Un)biased treatment}

\begin{table}
\begin{adjustbox}{width=\columnwidth}
\begin{tabular}{@{\extracolsep{5pt}}lccc}
\\[-1.8ex]\hline 
\hline \\[-1.8ex] 
 & \multicolumn{3}{c}{Perceived Fairness} \\ 
\cline{2-4} 
\\[-1.8ex] & Model 1 & Model 2 & Model 3 \\ 
\\[-1.8ex] & Coef.(S.E.) & Coef.(S.E.) & Coef.(S.E.)\\ 
\hline \\[-1.8ex] 
 Unfavorable Outcome \\ \hspace*{6mm} vs. Favorable Outcome & $-$1.040$^{***}$ (0.112) &  & $-$1.034$^{***}$ (0.111) \\ 
 Biased Treatment \\ \hspace*{6mm} vs. Unbiased Treatment &  & $-$0.410$^{***}$ (0.119) & $-$0.396$^{***}$ (0.111) \\ 
 Self-expected Pass \\ \hspace*{6mm} vs. Self-expected Fail & 0.682$^{***}$ (0.127) & 0.637$^{***}$ (0.135) & 0.655$^{***}$ (0.126) \\ 
  Constant & 4.063$^{***}$ (0.122) & 3.785$^{***}$ (0.133) & 4.278$^{***}$ (0.135) \\ 
 \hline \\[-1.8ex] 
R$^{2}$ & 0.164 & 0.059 & 0.182 \\ 
Adjusted R$^{2}$ & 0.161 & 0.055 & 0.177 \\ 
\hline 
\hline \\[-1.8ex] 
\textit{Note:}  & \multicolumn{3}{r}{$^{*}$p$<$0.05; $^{**}$p$<$0.01; $^{***}$p$<$0.001} \\ 
\end{tabular} 
\end{adjustbox}
\caption{\textbf{Regression models predicting perceived fairness from (un)favorable outcome to an individual (Model~1), (un)biased treatment across groups (Model~2), and both factors together (Model~3).}}
\label{tab:table1}
\end{table}

The models in Table~\ref{tab:table1} illustrate the impacts of (un)favorable outcome to individual and (un)biased treatment to group on the perception of fairness. Model 1 tests the effect of (un)favorable outcome to an individual, model 2 tests the effect of (un)biased treatment to a group, and model 3 includes both variables. All models include a control variable of self-expectation to control for the participants' own expectations of whether they would pass or fail.

\begin{figure}
    \centering
    \includegraphics[width=0.8\columnwidth]{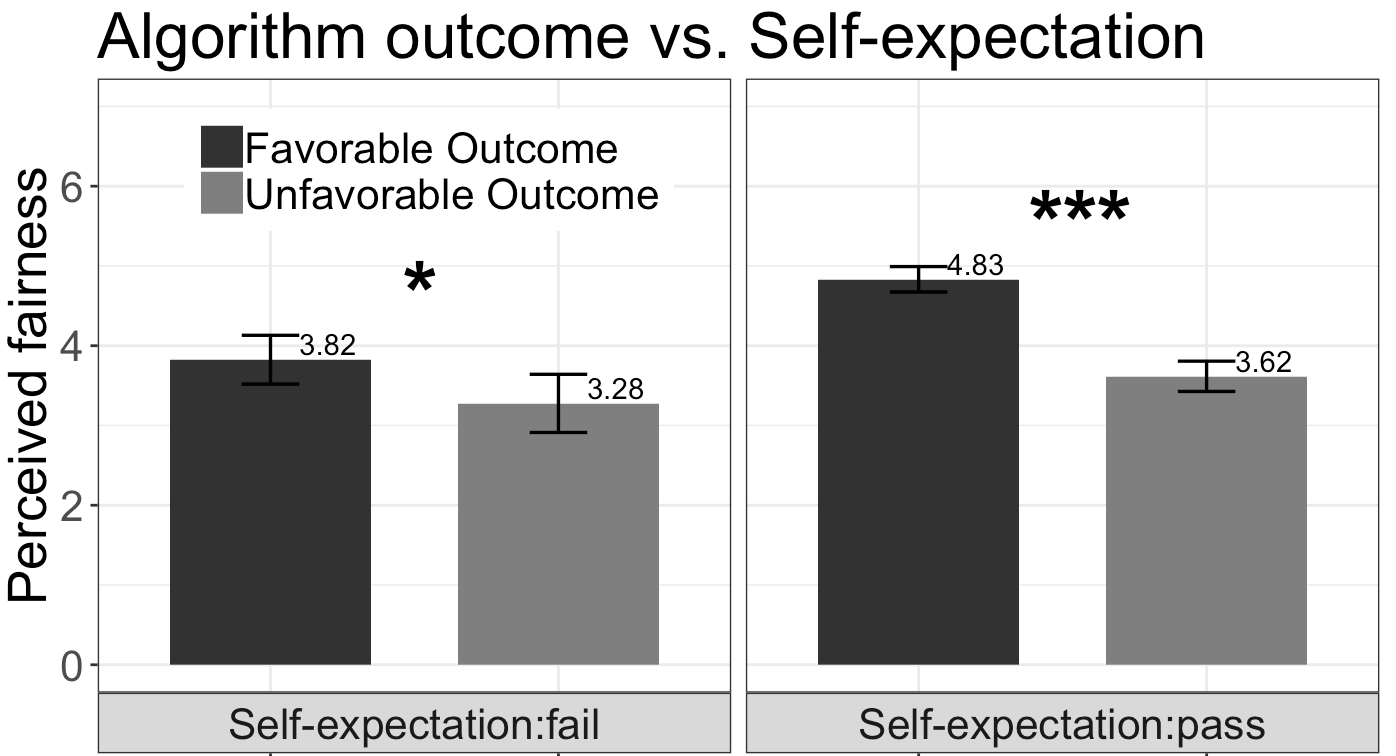}
    \caption{\textbf{The effects of algorithm outcome and self-expectation on perceived fairness. The negative effect of unfavorable outcome is significant no matter how the participants evaluated themselves. Error bars indicate 95\% confidence intervals.} $^{*}$p$<$0.05; $^{***}$p$<$0.001}
    \label{fig:outcomeSelf}
\end{figure}

Model 1 predicts that participants who were told that they failed the Master qualification test will rate the fairness of the algorithm 1.040 (95\% CI: [0.819, 1.260]) point lower on a 7-point scale, compared to participants who were told that they passed (p<0.001). Additional two-sample t-tests show that the difference of perceived fairness between receiving a favorable outcome and an unfavorable one is significant, both when participants expect themselves to fail the qualification algorithm (p=0.02) and when they expect to pass the qualification algorithm (p<0.001). Figure~\ref{fig:outcomeSelf} illustrates that this negative effect of unfavorable outcome is significant no matter how the users self-evaluated. Model 2 predicts that participants will rate the biased algorithm 0.410 (95\% CI: [0.175, 0.644]) points lower on a 7-point scale (p<0.01), compared with the unbiased algorithm. Therefore, both H1a and H1b are supported.

The effect of an unfavorable algorithm outcome is stronger than the effect of a biased algorithm. We compared the adjusted $R^2$ for models 1 and 2, finding that algorithm outcome explains more variance of the perceived fairness than whether the algorithm is biased or not (model 1 adjusted $R^2$=0.161, model 2 adjusted $R^2$=0.055). Model 3 compares the effect more explicitly. The effect size of (un)favorable outcome on perceived fairness (Coef.=-1.034, p<0.001, 95\% CI: [-1.253 -0.816]) is twice as much as the effect size of (un)biased algorithm (Coef.=-0.396, p<0.001, 95\% CI: [-0.615 -0.177]), which supports H1c.

\subsection{Algorithm Creation and Deployment} 

\begin{table}[h]
\centering 
 \begin{adjustbox}{width=\columnwidth}
\begin{tabular}{@{\extracolsep{5pt}}lcc} 
\\[-1.8ex]\hline 
\hline \\[-1.8ex] 
 & \multicolumn{2}{c}{Perceived Fairness} \\ 
\cline{2-3} 
\\[-1.8ex] & Model 1 & Model 2 \\ 
\\[-1.8ex] & Coef. (S.E.) & Coef. (S.E.)\\ 
\hline \\[-1.8ex] 
 Unfavorable Outcome vs. Favorable Outcome & $-$1.041$^{***}$ (0.113) & $-$0.887$^{***}$ (0.258) \\ 
  Biased Treatment vs. Unbiased Treatment & $-$0.395$^{***}$ (0.113) & $-$1.068$^{***}$ (0.257) \\ 
  CS Team vs. Outsourced & 0.131 (0.138) & $-$0.614$^{*}$ (0.240) \\ 
  CS and HR vs. Outsourced & 0.201 (0.137) & $-$0.352 (0.224) \\ 
  Machine Learning vs. Rules & 0.138 (0.112) & 0.086 (0.189) \\ 
  High Transparency vs. Low Transparency & $-$0.154 (0.114) & 0.056 (0.191) \\ 
  Mixed Decision vs. Algorithm-only & 0.090 (0.113) & 0.220 (0.189) \\ 
  Unfavorable Outcome {$\times$} CS team &  & 0.426 (0.272) \\ 
  Unfavorable Outcome {$\times$} CS and HR &  & 0.015 (0.273) \\ 
  Unfavorable Outcome {$\times$} Machine Learning &  & $-$0.238 (0.220) \\ 
  Unfavorable Outcome {$\times$} High Transparency &  & 0.106 (0.224) \\ 
  Unfavorable Outcome {$\times$} Mixed Decision &  & $-$0.420 (0.221) \\ 
  Biased Treatment {$\times$} CS team &  & 0.971$^{***}$ (0.272) \\ 
  Biased Treatment {$\times$} CS and HR &  & 1.070$^{***}$ (0.271) \\ 
  Biased Treatment {$\times$} Machine Learning &  & 0.343 (0.220) \\ 
  Biased Treatment {$\times$} High Transparency &  & $-$0.486$^{*}$ (0.224) \\ 
  Biased Treatment {$\times$} Mixed Decision &  & 0.174 (0.221) \\ 
  Self-expected Pass vs. Self-expected Fail & 0.659$^{***}$ (0.127) & 0.616$^{***}$ (0.125) \\ 
  Constant & 4.131$^{***}$ (0.178) & 4.438$^{***}$ (0.235) \\ 
 \hline \\[-1.8ex] 
R$^{2}$ & 0.190 & 0.235 \\ 
Adjusted R$^{2}$ & 0.178 & 0.210 \\ 
\hline 
\hline \\[-1.8ex] 
\textit{Note:}  & \multicolumn{2}{r}{$^{*}$p$<$0.05; $^{**}$p$<$0.01; $^{***}$p$<$0.001} \\ 
\end{tabular} 
\end{adjustbox}
\caption{\textbf{Regression models predicting perceived fairness from algorithm outcomes and development procedures. Model~1 shows the main effects, and Model~2 includes interaction terms.}} 
\label{fig:algdev} 
\end{table}

\begin{figure}[h]
  \centering
    \includegraphics[width=\columnwidth]{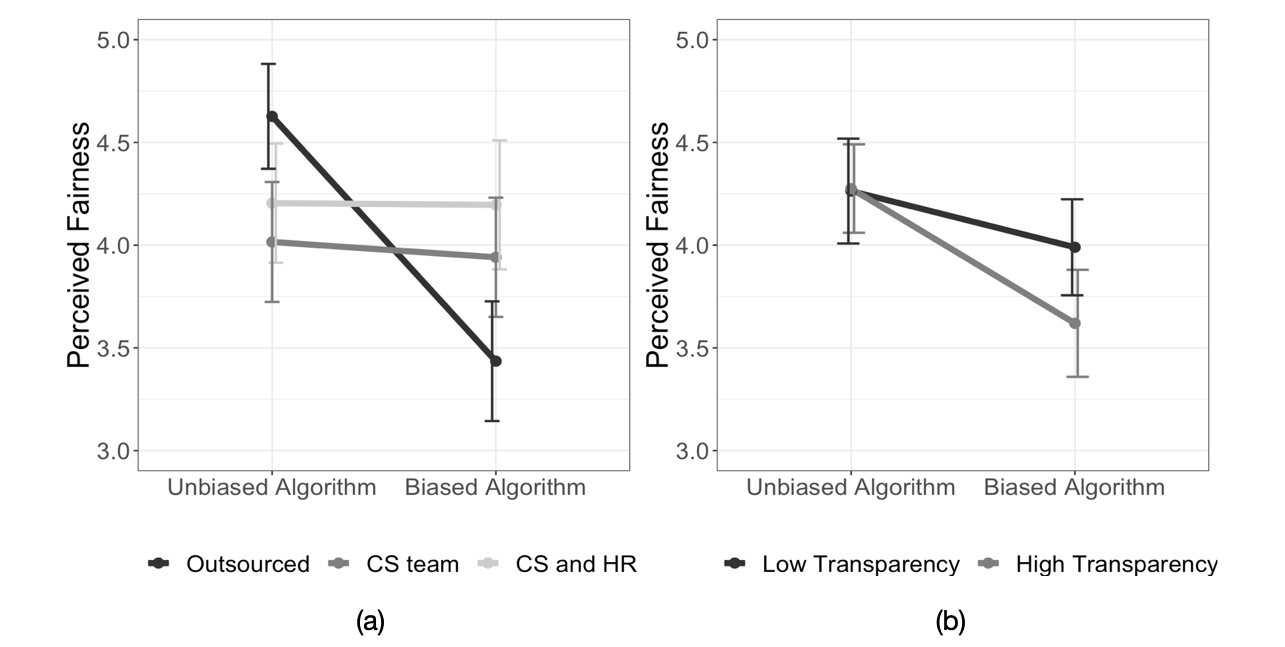}
    \caption{\textbf{The interaction between development procedures and algorithm outcomes on perceived fairness. Error bars indicate 95\% confidence intervals. The Y-axis represents the average rating of perceived fairness in the specific group. Figure (a) shows that describing a development procedure with ``outsourcing'' exacerbates the negative effect of biased treatments across groups. Figure (b) shows that higher level of transparency exacerbates the negative effect of biased treatments across groups.}}
    \label{fig:Al}
\end{figure}

Hypothesis 2 relates to how different algorithm creation and deployment procedures affect perceived fairness. Model 1 in Table~\ref{fig:algdev} examines the main effects of different algorithm development procedures on perceived fairness, which allows us to test hypothesis 2a and 2b. Model 2 in Table~\ref{fig:algdev} examines the interaction effects between development procedures and algorithm outcomes.

Model 1 in Table~\ref{fig:algdev} shows that the algorithm development manipulations had no significant main effects on perceived fairness. Therefore, Hypothesis 2a and 2b are not supported. The five variables reflecting the algorithm's development procedures collectively explained less than 1\% of the variance in perceived fairness. 

Although there is no significant main effect, Model 2 in Table~\ref{fig:algdev} finds several interaction effects between development procedures and algorithm outcomes. The effects of algorithm bias are moderated by the procedural factors of design and transparency. The negative effect of a biased algorithm is strongest when the algorithm is built by an outsourced team, as compared with when it is built by computer science experts within the organization (Coef.=0.971, p<0.001, 95\% CI: [0.437, 1.504]), or by a mix of computer science experts and other MTurk staff within the organization (Coef.=1.070, p<0.001, 95\% CI: [0.537, 1.602]). Figure~\ref{fig:Al} (a) visualizes this interaction effect. Transparency also exacerbates the effect of a biased algorithm on perceived fairness (Coef.=-0.486, p<0.05, 95\% CI: [-0.925, -0.046]), which is visualized in Figure~\ref{fig:Al} (b).

\subsection{Individual difference}

Table~\ref{fig:individual} describes a regression model predicting perceived fairness from individual differences (e.g., education level, computer literacy, and demographics). Model 1 describes  main effects, while Model 2 explores the interaction effects between these variables and algorithm outcomes.

\begin{table}[h]
\begin{adjustbox}{width=\columnwidth}
\begin{tabular}{@{\extracolsep{5pt}}lcc} 
\\[-1.8ex]\hline 
\hline \\[-1.8ex] 
 & \multicolumn{2}{c}{Perceived Fairness} \\ 
\cline{2-3} 
\\[-1.8ex] & Model 1 & Model 2 \\ 
\\[-1.8ex] & Coef. (S.E.) & Coef. (S.E.)\\ 
\hline \\[-1.8ex] 
 Unfavorable Outcome vs. Favorable Outcome & $-$1.057$^{***}$ (0.115) & $-$1.237 (0.651) \\ 
  Biased Treatment vs. Unbiased Treatment & $-$0.426$^{***}$ (0.114) & $-$1.070 (0.671) \\ 
  Low Literacy vs. High Literacy & $-$0.285$^{*}$ (0.117) & $-$0.384 (0.204) \\
  Above 45 vs. Between 25-45 & $-$0.142 (0.150) & 0.060 (0.271) \\ 
  Below 25 vs. Between 25-45 & 0.237 (0.208) & 0.069 (0.347) \\ 
  Male vs. Female & 0.209 (0.122) & $-$0.019 (0.203) \\ 
  Bachelor's Degree vs. Above Bachelor's Degree & $-$0.198 (0.188) & 0.070 (0.376) \\ 
  Below Bachelor's Degree vs. Above Bachelor's Degree & $-$0.272 (0.189) & 0.276 (0.371) \\ 
  Asian vs. Non-asian & 0.106 (0.282) & 0.031 (0.562) \\ 
  White vs. Non-white & $-$0.119 (0.220) & $-$0.534 (0.501) \\ 
  Black vs. Non-black & 0.340 (0.244) & $-$0.103 (0.544) \\ 
  Native American vs. Non-native American & 0.218 (0.390) & 0.690 (0.926) \\ 
  Hispanic vs. Non-hispanic & $-$0.136 (0.240) & $-$0.770 (0.524) \\ 
  Unfavorable Outcome {$\times$} Low Literacy &  & 0.268 (0.240) \\ 
  Unfavorable Outcome {$\times$} Above 45 &  & $-$0.334 (0.307) \\ 
  Unfavorable Outcome {$\times$} Below 25 &  & 0.215 (0.423) \\ 
  Unfavorable Outcome {$\times$} Male &  & 0.558$^{*}$ (0.242) \\ 
  Unfavorable Outcome {$\times$} Bachelor's Degree &  & $-$0.723 (0.389) \\ 
  Unfavorable Outcome {$\times$} Below Bachelor's Degree &  & $-$1.144$^{**}$ (0.393) \\ 
  Unfavorable Outcome {$\times$} Asian &  & 0.329 (0.624) \\ 
  Unfavorable Outcome {$\times$} White &  & 0.601 (0.533) \\ 
  Unfavorable Outcome {$\times$} Black &  & 0.918 (0.584) \\ 
  Unfavorable Outcome {$\times$} Native American &  & $-$0.600 (1.029) \\ 
  Unfavorable Outcome {$\times$} Hispanic &  & 0.117 (0.568) \\ 
  Biased Treatment {$\times$} Low Literacy &  & $-$0.034 (0.239) \\ 
  Biased Treatment {$\times$} Above 45 &  & $-$0.097 (0.306) \\ 
  Biased Treatment {$\times$} Below 25 &  & 0.135 (0.425) \\ 
  Biased Treatment {$\times$} Male &  & $-$0.097 (0.245) \\ 
  Biased Treatment {$\times$} Bachelor's Degree &  & 0.394 (0.387) \\ 
  Biased Treatment {$\times$} Below Bachelor's Degree &  & 0.181 (0.385) \\ 
  Biased Treatment {$\times$} Asian &  & 0.011 (0.631) \\ 
  Biased Treatment {$\times$} White &  & 0.452 (0.549) \\ 
  Biased Treatment {$\times$} Black &  & 0.063 (0.583) \\ 
  Biased Treatment {$\times$} Native American &  & $-$0.241 (1.030) \\ 
  Biased Treatment {$\times$} Hispanic &  & 1.234$^{*}$ (0.581) \\ 
  Self-expected Pass vs. Self-expected Fail & 0.598$^{***}$ (0.133) & 0.562$^{***}$ (0.137) \\ 
  Constant & 4.649$^{***}$ (0.319) & 4.877$^{***}$ (0.622) \\ 
 \hline \\[-1.8ex] 
R$^{2}$ & 0.220 & 0.262 \\ 
Adjusted R$^{2}$ & 0.200 & 0.209 \\ 
\hline 
\hline \\[-1.8ex] 
\textit{Note:}  & \multicolumn{2}{r}{$^{*}$p$<$0.05; $^{**}$p$<$0.01; $^{***}$p$<$0.001} \\ 
\end{tabular}  
\end{adjustbox}
  \caption{\textbf{Regression models predicting perceived fairness from algorithm outcomes and individual differences. Model~1 shows the main effects, and Model~2 includes interaction terms.}} 
  \label{fig:individual} 
\end{table} 

\begin{figure} [h]
   \centering
    \includegraphics[width=1\columnwidth]{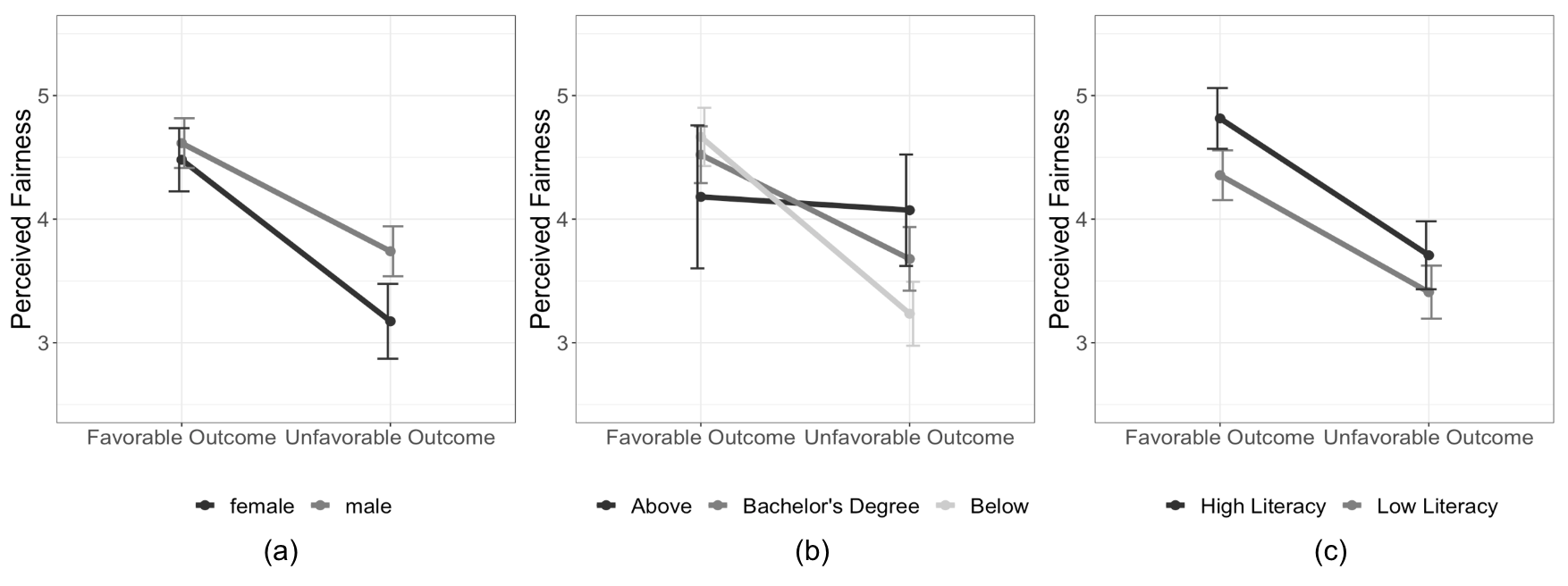}
    \caption{\textbf{Individual differences in the effects of (un)favorable outcome on perceived fairness. Error bars indicate 95\% confidence intervals. The Y-axis represents the average rating of perceived fairness in the specific group. (a) Gender: the effect of an unfavorable outcome is stronger for female participants. (b) Education Level: the effect of an unfavorable outcome is the smallest for participants with an education level greater than a Bachelor's degree. (c) Computer Literacy: participants with lower computer literacy tend to perceive the algorithm as less fair.}}
    \label{fig:demographic}
\end{figure}

Model 1 in Table~\ref{fig:individual} shows that computer literacy is positively correlated with perceived fairness. On average, participants with low computer literacy report significantly lower perceived fairness than participants with high computer literacy (Coef.=-0.285, p<0.05, 95\% CI: [-0.515, -0.055]), which supports hypothesis 3b. Figure~\ref{fig:demographic}~(c) visualizes the main effect of computer literacy. However, we do not observe a main effect of gender, education level, age, and race on the perceived fairness. Therefore, Hypothesis 3a and 3c are not supported and Hypothesis 3b is partially supported. 

Model 2 in Table~\ref{fig:individual} shows that certain groups react more strongly to an unfavorable outcome. For example, on average, female participants react more strongly to an unfavorable outcome than male participants (Coef.=0.558, p<0.05, 95\% CI: [ 0.082, 1.034]) and participants with a lower education level react more strongly to an unfavorable outcome than participants with a higher education level (Coef.=-1.144, p<0.01, 95\% CI: [-1.916, -0.372]). Figure~\ref{fig:demographic}~(a) and Figure~\ref{fig:demographic}~(b) visualize how gender and education level interact with the effects of unfavorable outcomes on perceived fairness.

\section{Discussion}

In this research, we seek to better understand the factors that impact perceptions of fairness around algorithmic decision-making, focusing on three sets of factors: algorithm outcomes, development procedures, and individual differences. 

\subsection{Algorithm Outcomes}

When a person is subjected to an algorithmic decision, the outcome of that decision is a very important factor in understanding how that person will evaluate the fairness of the process. In our study, participants who were told the algorithm gave them a ``fail'' decision rated the algorithmic decision-making process one point less fair, on average, on a seven point scale. The degree of this effect in our study was even greater than the effect of disclosing very biased prediction error rates across demographic groups, which caused participants to reduce their fairness evaluation by about 0.4 points (p<0.001). 

Before revealing the (randomly assigned) algorithm decision, we asked participants what decision they thought the algorithm would make. We find that the participants with the lowest overall aggregate fairness scores came from the group that had a ``fail'' self-expectation and an unfavorable outcome. This is surprising, since this group's expectations align with the algorithm's predictions, which we would intuitively think of as ``fair''. It is possible that people who think the algorithm will give them an unfavorable result generally have lower beliefs in the fairness of algorithmic decision-making, independent of the actual outcome. 
For example, in the open-ended response, one participant in the condition of ``biased treatment'' mentioned that she had ``fail'' self-expectation because she believed she would be disadvantaged by the system based on the description of the algorithm, independent of her judgement of her own quality of work: \textit{``The process as you all described is definitely not fair. Being a woman over 54 years of age, puts me at at least 2 disadvantages with this system.''}
In our sample, we find that participants whose expectation is ``fail'' rate the fairness of the process 0.7 points lower than participants whose expectation is ``pass'' (p<0.01).

\textit{\textbf{Implications.}} The important influence of an algorithm's outcome on our perception of its fairness has several possible implications. First, organizations seeking to build fair algorithmic decision-making processes must understand that fairness is complicated, and is a broader concept than just a measure of biases against particular groups. Even when the algorithm is presented as fair based on error rate measures, users' perceived fairness might still diverge due to the favorable or unfavorable outcome they receive personally. While recent work indicated that user feedback could be an important data source in evaluating algorithm fairness~\cite{holstein_improving_2019, eslami_user_2019}, our work suggests that systems that evaluate algorithm fairness through user feedback must measure or incorporate an ``outcome favorability'' bias in their models.

Another perspective is that we \textit{should} show algorithmic decisions to people to better understand if they are fair. The New York Times, in 2017 published an interactive survey to help people understand how a mathematical formula for determining legal immigration status to the United States would affect them personally~\cite{bui_how_2017}. While it is inevitable for algorithms to produce favorable outcomes to certain people and unfavorable outcomes to the other, it is possible that interactive tools like these help people put themselves in the shoes of those who would be affected by an algorithm directly, proving them with a different perspective on fairness and inducing empathy across users. We look forward to future work that seeks to develop interactive tools for understanding outcomes of complicated algorithmic processes on different types of people.

\subsection{Development Procedures}

The four manipulations varying the description of the algorithm's development process had no main effect on participants' fairness evaluations. We speculate that the lack of main effects may indicate a weak manipulation. It is possible that participants do not believe that the algorithms were actually developed as described. Another possibility is that participants, who are not algorithm experts, do not know to what extent ``how'' and ``who'' build the algorithms might impact the algorithm. 
For example, some participants believed more human involvement could make the decision-making process more fair. One participant wrote: \textit{``[The process] should have some human oversight to increase its fairness.''} However, other participants believed human involvement could bring in more potential biases. Another participant wrote: \textit{``I think it is somewhat fair because there is no human error from an Amazon employee.''} 
Non-expert stakeholders had different beliefs on how to build a ``fairer'' algorithmic system.  
The findings also suggest that the effect of development procedures is much smaller compared with the effect of algorithmic outcomes and biases to non-expert stakeholders.

Several interesting interaction effects emerge from these manipulations. In particular, both an ``outsourced'' process and a ``high transparency'' process exacerbate the negative effect of algorithmic bias on perceptions of fairness (see Figure~\ref{fig:Al}). One plausible explanation for the effect of outsourcing is that people feel an internal development team may have good reasons for biased error rates (e.g., to maximize overall accuracy), while an outsourced team may not have the MTurker's best interests in mind. Potentially, a more transparent process --- which includes publishing benchmarks about the process's accuracy --- triggers a feeling that the process \textit{should} be more fair, lowering the resulting fairness rating. Other prior research has also found complicated interactions between transparency and user perceptions of algorithms~\cite{poursabzi-sangdeh_manipulating_2018, kizilcec_how_2016}.

\textit{\textbf{Implications.}} Despite the possibility of weak experimental manipulation, one possible explanation of our results is that the relationship between development procedures and fairness is not immediately clear to lay audiences, pointing to further research on explanation tools that can bridge this gap.  Our findings also suggest that the effect of development procedures is much smaller compared with the effect of algorithmic outcomes and biases to non-expert stakeholders. Therefore, organizations seeking to build algorithmic systems that people perceive as fair should think beyond just development procedures, as these procedures in themselves do not necessarily have strong impacts on how lay audience think of the decision-making process as less or more fair. 

\subsection{Individual Differences}
In our experiment, people with low computer literacy provided lower fairness evaluations than people with high computer literacy. This finding aligns with~\cite{pierson_demographics_2017}, who found that an hour-long class session about algorithmic fairness caused students to more strongly favor the use of computer algorithms over human judges. 

Furthurmore, lower education levels appear to exacerbate the effect of getting a favorable or unfavorable outcome on perceived fairness. In our experiment, participants with lower education levels provide much higher fairness ratings when they receive a favorable outcome rather than an unfavorable outcome, while participants with the highest education level (more than a Bachelor's degree) barely change their fairness ratings as a result of the outcome (see Figure~\ref{fig:demographic}). 

\textit{\textbf{Implications.}} 
The results suggest that education can help people think beyond their own self-interest when considering the fairness of an algorithmic process that can affect many others. Therefore, developing algorithm-related education programs could help bring together people with different experience with algorithms and create a shared understanding towards the adoption of algorithmic decision-making.  

\subsection{Limitations \& Future Work}

We conducted this study on the Mechanical Turk platform, asking MTurkers to evaluate an algorithm for assigning the MTurk Master qualification. While this design has the advantage of asking participants about their perception of an algorithm in an area where they are personally invested and knowledgeable, there are a number of limitations. While we included an ``attention check'' question and five ``quiz'' questions to reinforce understanding of the decision-making process, some workers may have not read the scenario carefully, while others may have forgotten some details of the description by the time they evaluated fairness. 

To determine the effect of bias against demographic groups, we picked two fixed sets of values: one with very comparable error rates across groups, and the other with different error rates between groups. While we picked these bias rates based on earlier work~\cite{buolamwini_gender_2018}, we did not experiment with different values. Higher or lower error rates for different demographic groups may change the relative importance of bias versus other factors. Further, while we picked ``error rate'' to represent algorithmic bias, this is a simplistic representation, and especially keen participants may have wondered if the error rate was in favor of or against the groups with the higher error rates. For example, false positive rates and false negative rates have completely different implications for people who are subject to the algorithmic decision making. Future work could more carefully isolate different types of error rates between demographic groups. 

Generally speaking, the scenarios may be very influential in determining the judgments of fairness. It is possible that other scenarios (e.g., describing recidivism prediction, hiring, or school admission) would have led to very different results. We suggest that further work is necessary to understand the generalizability of these findings

\section{Conclusion}

In this research, we examine the factors that influence people's perceptions of the fairness of algorithmic decision-making processes using a between-subjects randomized survey on Mechanical Turk. We focus our manipulations and analysis on several factors: algorithm outcomes, development procedures, and individual differences. We find that people's evaluations of fairness are very sensitive to whether or not they receive a positive outcome personally, even surpassing the negative effect of describing an algorithm with strong biases against particular demographic groups. 

\section{Acknowledgments}
We want to thank Zhiwei Steven Wu, Hao-Fei Cheng, and members of the GroupLens Lab for their help in this project. This work was supported by the National Science Foundation (NSF) under Award No. IIS-2001851 and No. IIS-2000782, and the NSF Program on Fairness in AI in collaboration with Amazon under Award No. IIS-1939606.

%
%
%
%
%
\balance{}

\balance{}

\bibliographystyle{SIGCHI-Reference-Format}
\bibliography{sample}

\end{document}